\pdfoutput=1
\documentclass[11pt,a4paper]{article}
\usepackage{jcappub}
\allowdisplaybreaks
\usepackage{xcolor,url}
\usepackage{amsmath}
\usepackage{graphicx}
\usepackage{esint}
\usepackage{slashed}
\usepackage{bigints}
\usepackage[english]{babel}
\usepackage{comment}
\usepackage{hyperref}
\usepackage{multirow}

\usepackage{babel}
\begin{document}

\title{Stellar limits on light CP-even scalar}

\author[a]{P. S. Bhupal Dev,}
\author[b]{Rabindra N. Mohapatra,}
\author[c,a]{Yongchao Zhang}

\affiliation[a]{Department of Physics and McDonnell Center for the Space Sciences,  Washington University,
St. Louis, MO 63130, USA}

\affiliation[b]{Maryland Center for Fundamental Physics, Department of Physics, University of Maryland,
College Park, MD 20742, USA}

\affiliation[c]{School of Physics, Southeast University, Nanjing 211189, China}

\emailAdd{ bdev@wustl.edu, rmohapat@umd.edu, zhangyongchao@seu.edu.cn}

\abstract{We revisit the astrophysical constraints on a generic light CP-even scalar particle $S$, mixing with the Standard Model (SM) Higgs boson,  from observed luminosities of the Sun, red giants, {white dwarfs} and horizontal-branch stars. The production of $S$ in the stellar core is dominated by the electron-nuclei bremsstrahlung process $e + N \to e + N + S$.  With the $S$ decay and reabsorption processes taken into consideration, we find that the stellar luminosity limits exclude a broad range of parameter space in the $S$ mass-mixing plane, with the scalar mass up to 350 keV and the mixing angle ranging from {$7.0\times10^{-18}$} to $3.4\times10^{-3}$.  We also apply the stellar limits to a real-singlet scalar extension of the SM, where we can relate the mixing angle to the parameters in the scalar potential. In both the generic scalar case and the real-singlet extension, we show that the stellar limits preclude the scalar interpretation of the recently observed XENON1T excess in terms of the $S$ particles emitted from the Sun.
}

%\keywords{Stars, Real Scalar}

\maketitle
%\tableofcontents

\section{Introduction}

The astrophysical considerations based on the observed luminosities of stellar objects such as the Sun, red giants (RGs), {white dwarfs (WDs)}, and  horizontal-branch (HB) stars can impose stringent constraints on light beyond the Standard Model (BSM) particles coupling to the SM sector~\cite{Raffelt:1996wa}. If the masses of the new BSM particles are in the (sub-)keV range, the core temperature of the stellar objects is large enough to enable their production inside the core. Once produced, depending on the rate of production (or the coupling to the SM particles), they can lead to additional energy loss mechanisms beyond the standard photon emission. Demanding the extra energy loss rate to be less than the observed luminosities then leads to bounds on the couplings of the new particles to SM particles. On the other hand, if the couplings to SM particles become too large, the particle may scatter strongly against the SM particles (electrons, nucleons or photons) or be reabsorbed inside the stellar core to have a mean free path (MFP) which is less than the size of the stellar objects. In this case, the BSM particle will be emitted only with a thermal rate and will not lead to uncontrolled energy loss, thereby allowing such large couplings by the luminosity constraints. Using these arguments, stellar limits on the masses and couplings of various light BSM particles, such as scalars, pseudoscalars, and dark photons, have been derived~\cite{Sato:1975vy, Dicus:1978fp, Sato:1978vy, Ellis:1982ej, Iwamoto:1984ir, Pantziris:1986dc, Frieman:1987ui, Grifols:1988fv, Blinnikov:1990ui,  An:2013yfc, Redondo:2013lna, Hardy:2016kme}. Similar astrophysical limits using the inferred supernova luminosity of SN1987A have also been derived~\cite{Turner:1987by, Raffelt:1987yt, Mayle:1987as, Brinkmann:1988vi,  Burrows:1988ah,  Ishizuka:1989ts, Arndt:2002yg, Dent:2012mx, Kazanas:2014mca, Krnjaic:2015mbs, Chang:2016ntp, Lee:2018lcj, Dev:2020eam}, which extend to higher masses $\sim {\cal O}(100~{\rm MeV})$ of the BSM particles, because of the higher supernova core temperature; however, for the (sub-)keV-scale BSM particles of interest here, the stellar limits usually turn out to be more stringent than the supernova limits.

In this paper, we revisit the stellar constraints for a generic light CP-even scalar $S$, couplings to the SM only via an effective mixing angle $\sin\theta$ with the SM Higgs boson $h$. We find that the scalar production via the bremsstrahlung process $e + N \to e + N + S$ is dominant over other production channels, such as via Compton-like, Primakoff-like, $NN/ee$ bremsstrahlung and {plasma mixing} processes inside the stellar core {(cf.~Fig.~\ref{fig:channels}).}  We also take into account the decay of $S$ into photons, as well as the reabsorption of $S$ inside the core via the inverse bremsstrahlung process $e + N+S \to e + N$ to calculate the MFP of $S$.  We then use the luminosity limits of the Sun, RGs, {WDs} and HB stars to derive updated stellar constraints on the scalar mass $m_S$ and the mixing angle $\sin\theta$. These stellar limits are found to be much more stringent than those from supernovae~\cite{Dev:2020eam}, invisible meson decays~\cite{Dev:2017dui,Dev:2019hho}, and the LHC limits on invisible decay of the SM Higgs~\cite{Sirunyan:2018owy, ATLAS:2020cjb}, for scalar masses below roughly 200 keV. Our main results are shown in Fig.~\ref{fig:limits}.

The stellar limits for a generic scalar obtained in this paper can be easily adapted to more specific or ultraviolet-complete models. As an explicit example, we consider a real-singlet scalar extension of the SM, where we can relate the $h-S$ mixing angle $\theta$ to the scalar quartic coupling with the SM Higgs. This model is of particular interest, as for sufficiently small mixing, the light scalar could serve as the dark matter (DM) candidate. We find that the {WD limits derived here impose a lower bound of $m_S\gtrsim 10$ keV on the parameter space required for successful DM relic density, as shown in Fig.~\ref{fig:limits:singlet}.}

%~(ii) a leptonic scalar, whose production in the stars will be predominantly induced by its coupling to electrons, instead of nucleons.
%The stellar limits on the singlet and leptonic scalars are shown respectively in Figs.~\ref{fig:limits:singlet} and \ref{fig:limits:leptonic}.
%For completeness, we also investigate the supernova limits on the leptonic scalar $S$, which turns out to be very different from the generic case recently studied in Ref.~\cite{Dev:2020eam}.

%couples directly to SM fermions as may be the case in some multi-Higgs models or majoron models.
%Except for the first case, we do not discuss the detailed model origin of the scalar field.

%We also discuss the supernova constraints for the second case where the SM scalar couples directly to SM leptons.
%The SN constraints on the first case was discussed by us in a recent paper.

We then explore whether  abundant production of $S$ in the Sun could provide an explanation of the recently observed XENON1T excess at the keV-scale~\cite{Aprile:2020tmw}, without violating the stellar emission bounds. This is conceptually similar to the solar axion interpretation of the XENON1T excess~\cite{Aprile:2020tmw, DiLuzio:2020jjp, Gao:2020wer, Dent:2020jhf,  Sun:2020iim, Li:2020naa, Athron:2020maw}.\footnote{The XENON1T excess is found to be compatible with a recent search of solar axions by PandaX-II~\cite{Zhou:2020bvf}.} However, as in the axion case, it turns out that, for both the generic case and the singlet model we consider, the stellar luminosity constraints seem to preclude the possibility of fitting the XENON1T excess using the $S$ particles emitted from the Sun.
%See for example Figs.~\ref{fig:excess} and \ref{fig:excess:leptonic}.
It should however be noted that the stellar constraints discussed here can in principle be avoided by adding new interactions for the BSM  particles {that suppresses single $S $ production by changing the nature of phase transition~\cite{Mohapatra:2006pv}, mixing with other particles~\cite{Masso:2006gc, Brax:2007ak, Bloch:2020uzh, Budnik:2020nwz}, or using interactions that make them more massive in a matter-rich environment~\cite{DeRocco:2020xdt}.} However, we do not invoke any such exotic mechanisms here.

The rest of this paper is organized as follows: In Section~\ref{sec:channels}, we consider the generic model where the $S$ field interacts with the SM fields via an effective mixing with the SM Higgs field and then lay out the different mechanisms for its production in various astrophysical sites. In Section~\ref{sec:limits}, we obtain the constraints on the scalar mass $m_S$ and the $h-S$ mixing angle $\sin\theta$ from the stellar luminosity constraints. We then consider in Section~\ref{sec:excess} the possibility of the $S$ particle emitted from the Sun fitting the XENON1T excess. In Section~\ref{sec:singlet} we apply the stellar luminosity limits to a real singlet-scalar extension of the SM, where the mixing with SM Higgs is caused by the vacuum expectation value (VEV) of the scalar field $S$, and discuss its phenomenological implications. We conclude in Section~\ref{sec:conclusion}. Some of the differential cross sections relevant for $S$ production are collected in Appendix~\ref{sec:differential}.

%\section{Framework of effective mixing with the SM Higgs}
%\label{sec:generic}

\section{Production channels}
\label{sec:channels}

In this section, we work in a model-independent way, assuming that all the couplings of the light scalar $S$ to the SM particles are from its mixing with the SM Higgs, without regard to how the mixing arises. As a result, there are only two parameters which play a role in our discussion, i.e. the scalar mass $m_S$ and the mixing angle $\sin\theta$ with the SM Higgs.
%\subsection{Production channels}
Through this mixing, the scalar $S$ couples to electrons, nucleons and pions at the tree-level and to photons at the one-loop level. Then $S$ can be produced inside the stars in the following channels:%\footnote{Here we ignore the plasma mixing effects~\cite{Hardy:2016kme}, in which the light scalar can mix with the SM plasmons -- the in-medium longitudinal mode of the photon, and can be resonantly produced down to arbitrarily low masses.}
\begin{itemize}
    \item {\it Compton-like}: %$e\gamma \to e S$,
    \begin{eqnarray}
    \label{eqn:Compton}
    e+ \gamma \to e + S \,,
    \end{eqnarray}
    which is induced by the coupling of $S$ to electrons.
    \item {\it Primakoff-like}: %$\gamma + X \to X+ S$ (with $X$ being electron or nuclei),
    \begin{eqnarray}
    \label{eqn:Primakoff}
    \gamma + X \to X+ S \text{ (with $X$ being electrons or nuclei)} \,,
    \end{eqnarray}
    which is induced by the loop-level coupling of $S$ to photons.

     \item {\it $e-N$ bremsstrahlung}: %$eN\to eNS$, with $N = p$ or nuclei,
    \begin{eqnarray}
    \label{eqn:brem2}
    e + N \to e + N + S  \text{ (with $N$ being nuclei)} \,,
    \end{eqnarray}
    which is mediated by a photon, and $S$ couples predominantly to $N$. %Note that if $S$ is a leptonic scalar, its coupling to electrons will be much larger than that to nucleus; see Section~\ref{sec:leptonic} for more details.

    \item {\it $N-N$ bremsstrahlung}: %$N + N \to N + N S$ (with $N = p$ or nuclei),
    \begin{eqnarray}
    \label{eqn:brem1}
    N + N \to N + N + S \text{ (with $N$ being nuclei)} \,,
    \end{eqnarray}
    which is mediated by either a pion or a photon.

    \item {\it $e-e$ bremsstrahlung}:
    \begin{eqnarray}
    \label{eqn:brem3}
    e + e \to e + e + S \,,
    \end{eqnarray}
    which is mediated by a photon.

    \item {{\it Plasma mixing effect}: Inside the stellar core with large electron density and thermalized plasma, the light scalar can mix with the plasmon, i.e.~the in-medium longitudinal mode of the SM photon. As a result, when the scalar mass is smaller than the plasma frequency, the scalar can be produced resonantly via the collective plasma excitation. }
\end{itemize}

For the production of $S$ via the Compton-like process $e+\gamma \to e + S$, the total cross section is given by~\cite{Araki:1950} (see Appendix~\ref{sec:differential} for the differential cross section with respect to the $S$ energy $E_S$)
\begin{eqnarray}
\label{eqn:Compton:total}
\sigma_{\rm C} &=& \frac{\alpha y_e^2 \sin^2\theta}{3m_e^2}
f_{\rm C}(q,y) \, .
\end{eqnarray}
Here the subscript ``C'' stands for ``Compton'', $\alpha=e^2/4\pi$ is the fine-structure constant, $y_e$ is the electron Yukawa coupling in the SM, $m_e$ is the electron mass, $q \equiv m_S/m_e$,
%$x \equiv E_S/m_e$ (with $E_S$ the energy of $S$),
$y \equiv E_\gamma/m_e$ (with $E_\gamma$ being the photon energy), and the function $f_{\rm C} (q,y)$ is defined in Eq.~\eqref{eqn:fC}. As the electron mass is much larger than the keV-scale stellar core temperatures,
%(in supernovae it is very different, see Section~\ref{sec:supernova} for more details),
for simplicity we have neglected the kinetic energy of electrons and assumed the electrons to be at rest in the initial state. The scalar emission rate per unit volume is then given by
\begin{eqnarray}
\label{eqn:rate:compton}
Q_{\rm C} \simeq n_e \int \frac{2 d^3 {\bf k}_\gamma}{(2\pi)^3} \frac{E_\gamma}{e^{E_\gamma/T}-1}
\sigma_{\rm C}  \,,
\end{eqnarray}
where ${\bf k}_\gamma$ is the photon 3-momentum with $E_\gamma=|{\bf k}_\gamma|$, $n_e$ is the number density of electrons and $T$ is the temperature in the stellar core. In Eq.~(\ref{eqn:rate:compton}) we have used the approximation that $E_\gamma \simeq E_S$.

For the Primakoff process in Eq.~(\ref{eqn:Primakoff}), the coherent production cross section is~\cite{Dicus:1978fp}
\begin{eqnarray}
\label{eqn:xs:Primakoff}
\sigma_{{\rm P},\, X} =  64\pi Z_X^2 \alpha  \frac{E_\gamma \Gamma (S\to \gamma\gamma)}{m_S^2}
\frac{\sqrt{E_\gamma^2 - m_S^2} (E_\gamma - m_S)}{(m_S^2 + 2 m_S E_\gamma + k_{\rm scr}^2)^2} \, .
\end{eqnarray}
Here the subscript ``P'' stands for ``Primakoff'', $Z_X$ is the atomic number of the nucleus $X$ (for electrons, we can set $Z_X=1$), and the scalar decay width~\cite{Dev:2017dui}\footnote{For $m_S< 2m_e$, the diphoton channel is the only dominant decay mode of $S$. Other possible decay modes like $S\to \nu\bar{\nu}$ (via $Z$ loop), $S\to \gamma Z^*\to \gamma \nu\bar{\nu}$ and $S\to Z^* Z^*\to 4\nu$ have orders of magnitude smaller partial widths and thus can be safely neglected.}
\begin{eqnarray}
\label{eqn:width}
\Gamma (S \to \gamma\gamma) = \frac{\alpha^2 m_S^3 \sin^2\theta}{512 \pi^3 v_{\rm EW}^2} \frac{121}{9} \,,
\end{eqnarray}
with the factor of $121/9$ from summing up the loop factors for all the charged SM particles in the limit of $m_S \to 0$ and $v_{\rm EW}$ is the electroweak VEV.  In the limit of massless $S$, the cross section in Eq.~(\ref{eqn:xs:Primakoff}) is divergent, therefore we have introduced the screening scale in the propagator~\cite{Raffelt:1996wa}
%[See Eq. (5.7) of~\cite{Raffelt:1996wa}]
\begin{eqnarray}
k^2_{\rm scr} = \frac{4\pi\alpha}{T} n_B \left(Y_e + \sum_j Z_j^2 Y_j \right) \,,
\end{eqnarray}
where $Y_{e,\,j}$ are the number fractions of electrons and the baryons $j$, and $n_B$ is the baryon number density. Then the energy loss rate per unit volume for the Primakoff process is
\begin{eqnarray}
\label{eqn:rate:Primakoff}
Q_{\rm P} = \sum_X n_X \int \frac{2 d^3 {\bf k}_\gamma}{(2\pi)^3} \frac{1}{e^{E_\gamma/T}-1} E_S \sigma_{{\rm P},\, X} \, ,
\end{eqnarray}
where $n_X$ is the number density of the corresponding $X$ particle.

For the bremsstrahlung processes, let us first consider the $e-N$ channel in Eq.~(\ref{eqn:brem2}), which is mediated by a photon and $S$ is emitted from the $N$ lines. The energy emission rate per unit volume in the star is given by
\begin{align}
\label{eqn:rate:master}
Q_{\rm B}^{(eN)} & = & \sum_i
\int {\rm d} \Pi_5 \sum_{\rm spins} |{\cal M}_i|^2 (2\pi)^4
\delta^4 (p_1 + p_2 - p_3 - p_4 - k_S) E_S f_1^{(e)} f_2^{(N_i)}  \, .
\end{align}
Here the subscript ``B'' denotes ``bremsstrahlung'', ${\rm d} \Pi_5$ is the $2\to 3$ phase space factor, $\mathcal{M}_i$'s are the coherent scattering amplitudes for the nuclei $N_i$, $p_{1,\, 2}$ and $p_{3,\,4}$ are the momenta of $e$ and $N_i$ in the initial and final states respectively, $k_S$ is the outgoing momentum of $S$, and $f_{1,\,2}$ are the non-relativistic Maxwell-Boltzmann distributions of the incoming electron and nucleons in the non-degenerate limit, defined by
\begin{eqnarray}
f^{(X)} = \frac{n_{X}}{2}
\left( \frac{2\pi}{m_{X} T} \right)^{3/2}
\exp \left\{ - \frac{{\bf p}^2_{X}}{2m_{X} T}\right\} \,,
\end{eqnarray}
with ${\bf p}_X$  being the 3-momentum of $X$. Following the calculations in Ref.~\cite{Dev:2020eam}, the emission rate in Eq.~(\ref{eqn:rate:master})  can be simplified as
\begin{eqnarray}
\label{eqn:QB:eN}
Q_{\rm B}^{(eN)} &=& \Big( \sum_i Z_{N_i}^2 A_{N_i}^2 n_{N_i} \Big)
\frac{ \alpha^2 y_{N}^2 \sin^2\theta T^{1/2} n_e}{\pi^{3/2} m_e^{3/2}} \int_{q}^{\infty} du \int_0^\infty dv \int_q^\infty dx \int_{-1}^1 dz \nonumber \\
&& \qquad  \times  \sqrt{uv} e^{-u} \sqrt{x^2 - q^2} \frac{\delta (u-v-x)}{(u+v-2\sqrt{uv}z)^2}  \,,
\end{eqnarray}
with the dimensionless parameters defined as
\begin{align}
\label{eqn:uvxyqz}
& u \ \equiv \ \frac{({\bf p}_1 - {\bf p}_2)^2}{m_N T} \,, \qquad
 v \ \equiv \ \frac{({\bf p}_3 - {\bf p}_4)^2}{m_N T} \,, \qquad
 x \ \equiv \ \frac{E_S}{T} \,, \nonumber  \\
& q \ \equiv \ \frac{m_S}{T} \,, \qquad
 z \ \equiv \ \cos(\theta_{if}) \,,
\end{align}
with ${\bf p}_{1,\,2,\,3,\,4}$ the 3-momenta of electrons and nucleons, and $z$ the angle between ${\bf p}_1 - {\bf p}_2$ and ${\bf p}_3 - {\bf p}_4$.
In Eq.~(\ref{eqn:QB:eN}) we have summed up the coherent contributions from all the nuclei elements $N_i$, with $n_{N_i}$ being the corresponding number density in the stars, $Z_{N_i}$ and $A_{N_i}$ being the atomic and mass numbers of $N_i$ respectively,\footnote{These factors come from the coherent couplings of the photon to protons and $S$ to nucleons respectively~\cite{Blinnikov:1990ui}, because the nuclear binding energies are well above the stellar core temperatures. This is different from the supernova core, where the coherence is lost due to higher core temperatures.} and $y_N\sim 10^{-3}$ the effective coupling of SM Higgs to nucleons~\cite{Shifman:1978zn, Cheng:1988im}. For simplicity we have taken the leading-order approximation that the couplings of SM Higgs to protons and neutrons are equal. {It is important to note here that there is no nucleon-mass suppression in Eq.~\eqref{eqn:QB:eN}, because the $m_N$ factors from the matrix element, phase space and distribution function coincidentally cancel each other.}

%(for the coupling of $S$ to $N$)
%\blue{(note that here we have the coherence effect for $Z,\, A >1$)}.

Regarding the $N-N$ bremsstrahlung process in Eq.~(\ref{eqn:brem1}), it can be mediated either by a pion or by a photon. For the pion-mediated channel, we compare it with the energy loss rate per unit volume in the $e -N $ channel in Eq.~(\ref{eqn:QB:eN}), which turns out to be:
%Comparing the $ep$ and $pp$ bremsstrahlung processes, which are mediated by photon and pion,
\begin{eqnarray}
\label{eqn:comparison}
\frac{Q_{\rm B}^{(NN)}}{Q_{\rm B}^{(eN)}} \sim \frac{(2m_N/m_\pi)^4 A_N^4 f_{pp}^4}{e^4} \frac{m_e^2}{m_N^2} \frac{T^4}{m_\pi^4} \frac{m_S^2}{m_N^2} \ll 1 \, .
\end{eqnarray}
Here $m_\pi$ and $m_N$ are respectively the masses of pions and nuclei, and $f_{pp}\simeq 1$ is the coupling of pion to protons. The first three ratios in Eq.~(\ref{eqn:comparison}) are respectively from the couplings, the phase space, the propagator, and the last one is due to the cancellation effect for the CP-even scalar in the $N-N$ channel~\cite{Dev:2020eam}. As can be seen from Eq.~(\ref{eqn:comparison}), the energy loss due to pion-mediated $N-N$ scattering process is much smaller than that for the $e-N$ channel. For the photon-mediated $N-N$ bremsstrahlung process, it is still suppressed by the phase space factor of $m_e^2/m_N^2$, with respect to the $e-N$ bremsstrahlung. Therefore, we will neglect the $N-N$ bremsstrahlung process in the following sections. Similarly, for the $e-e$ bremsstrahlung process in Eq.~(\ref{eqn:brem3}), the energy emission rate $Q_{\rm B}^{(ee)}$ will be highly suppressed as a result of the small electron Yukawa coupling $y_e$ in the SM. As a result, the bremsstrahlung production of $S$ will be dominated by the $e-N$ channel, with the energy loss rate per unit volume given in Eq.~(\ref{eqn:QB:eN}), which we will now apply to various astrophysical objects, such as the Sun, RGs, {WDs} and HB stars.

{As for the resonant plasma contribution, this is dominated by the coupling of $S$ to electrons, as the nucleon contribution is suppressed by ${\cal O}(m_e^2/m_N^2)$ relative to the plasma frequency $\omega_p$. For the generic scalar case discussed here and for $m_S<\omega_p<T$, the resonant production rate is given by~\cite{Hardy:2016kme}
\begin{eqnarray}
\label{eqn:Qplm}
Q_{\rm pl} \ \simeq \ \frac{y_e^2\sin^2\theta} {\alpha}k^2_{\omega_p}\omega_p^3
\frac{1}{e^{\omega_p/T}-1} \,,
\end{eqnarray}
where the  subscript ``pl'' stands for ``plasma'' and $k^2_{\omega_p}=E_S^2-m_S^2$ is the squared 3-momentum of $S$ evaluated at energy $E_S=\omega_p\simeq \sqrt{n_e e^2/m_e}$.
%Here for simplicity we do not include any dependence of $Q_{\rm plm}$ on the scalar mass; see e.g.~Ref.~\cite{Guarini:2020hps} for more details.
The plasma effect could also modify to some extent all the other production rates discussed above; however, these thermal corrections are expected to be subdominant in stellar cores~\cite{Hardy:2016kme}.}

Given the new energy loss rates $Q$ in Eqs.~\eqref{eqn:rate:compton}, \eqref{eqn:rate:Primakoff}, {\eqref{eqn:QB:eN} and \eqref{eqn:Qplm}}, we can compute the corresponding luminosities as
\begin{align}
{\cal L}_S = VQP_{\rm decay} P_{\rm abs} \, ,
\label{eqn:lumi}
\end{align}
where $V=\frac{4\pi}{3}R^3$ is the volume of the stellar core,
\begin{eqnarray}
P_{\rm decay}^{} = \exp \left[- \frac{m_S}{E_S}  R \Gamma (S\to \gamma\gamma) \right]
\label{eqn:Pdecay}
\end{eqnarray}
is the decay probability factor (as only the $S$ particles decaying outside the star will contribute effectively to energy loss), with $m_S/E_S$ being the inverse Lorentz boost factor and $R$ being the radius of the stellar core, and
\begin{eqnarray}
P_{\rm abs} = \exp \left[ -  \frac{R}{ \langle \lambda_{} \rangle } \right]
\label{eqn:Pabs}
\end{eqnarray}
is the probability factor to account for the reabsorption of $S$ inside the star, with $\langle \lambda \rangle$ being the energy-averaged MFP of $S$ (cf.~Eq.~\eqref{eqn:MFP3}).
\begin{figure*}[t!]
    \centering
    \includegraphics[width=0.55\textwidth]{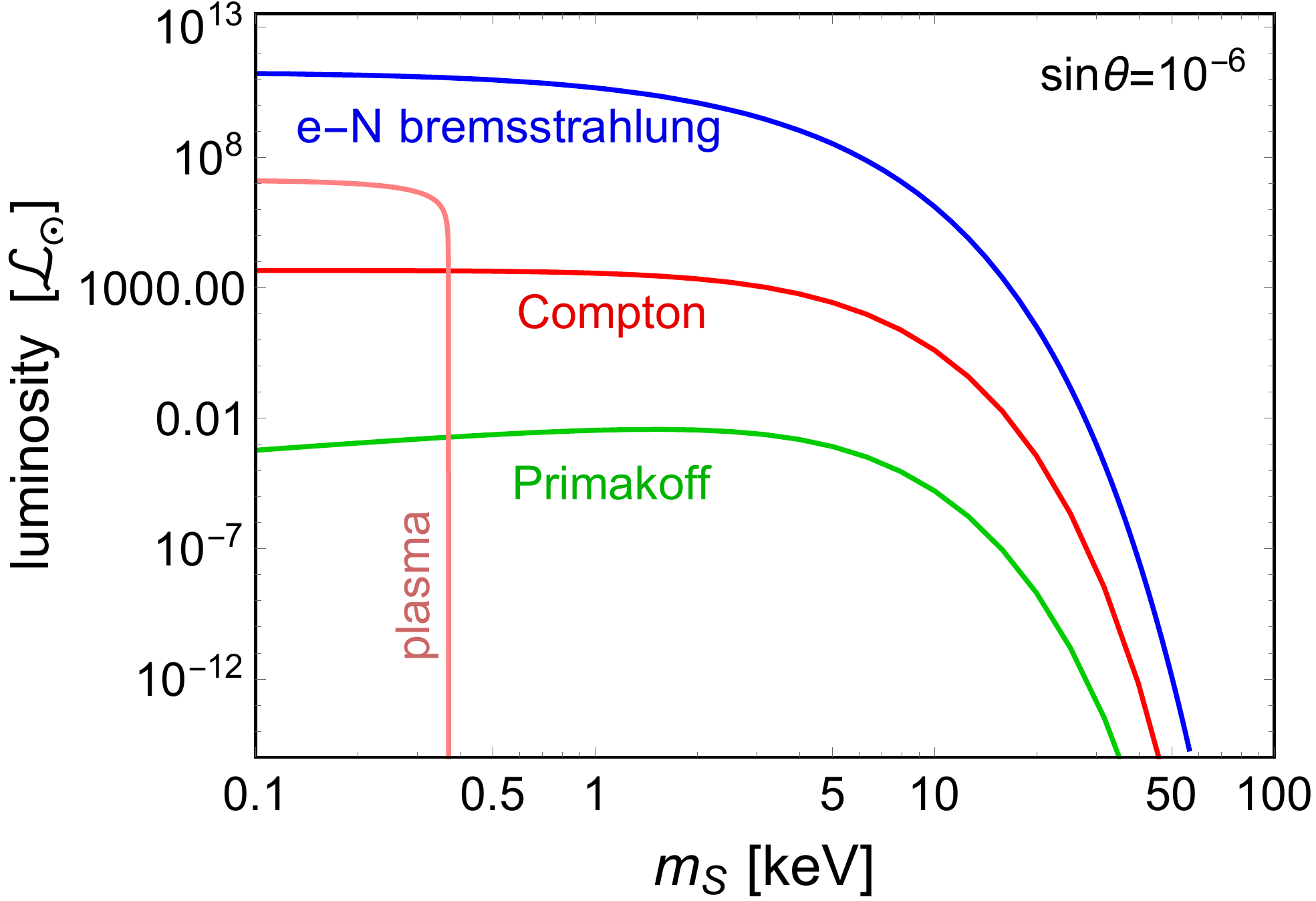}
    \caption{Comparison of the new solar luminosities due to the production of $S$ via the Compton-like (red), Primakoff-like (green), $e$-$N$ bremsstrahlung (blue)  processes, {and the plasma effect (pink)}. The mixing angle $\sin\theta$ is fixed at $10^{-6}$.}
    \label{fig:channels}
\end{figure*}
For illustration, taking a benchmark  value of $\sin\theta = 10^{-6}$, the solar luminosity $VQ$ (without the decay and absorption factors) in the {four} new channels mentioned above are shown in Fig.~\ref{fig:channels} as a function of the scalar mass.  For other values of $\sin\theta$, the luminosity limits can be scaled as $\sin^2\theta$. Here we have used the fact that the solar core consists of about 75\% of Hydrogen and 25\% of $^4$He, the core temperature $T_\odot\sim 1$ keV, the electron number density $n_e \simeq 10^{26} \, {\rm cm}^{-3}$, and the core size $R_\odot \simeq 7 \times10^{10}$ cm (see Table~\ref{tab:stars}). As expected, the luminosity in the $e-N$ bremsstrahlung channel is much larger than that from the Compton-like process, due to the larger Yukawa coupling $y_N$. The Primakoff channel, on the contrary, is highly suppressed due to the loop-level coupling of $S$ to photons. It is also clear from Fig.~\ref{fig:channels} that all these production channels are Boltzmann-suppressed for $m_S\gtrsim T_\odot$. {The plasma effect is only significant for the scalar mass below $\omega_p \simeq 0.37$ keV in the Sun, but still subleading to the $e-N$ bremsstrahlung process}.\footnote{{One should note that if the scalar $S$ is a leptonic scalar~\cite{Dev:2017ftk}, the couplings to nucleons will be absent at the tree-level, and can only arise at one-loop level from the leptonic couplings and the SM Higgs mediator. In this case, the plasma effect will be the dominant production mode if the scalar mass $m_S \ll \omega_p$~\cite{Hardy:2016kme}.}  }
%\blue{Here for simplicity we have neglected the dependence of plasma process on the scalar mass but take a sharp cutoff at $\omega_p$. }

Comparing the new luminosities~\eqref{eqn:lumi} with the observed stellar luminosities, we can put constraints on the scalar mass and mixing, as discussed in Section~\ref{sec:limits}. Before proceeding to the stellar limits, we would like to emphasize that the CP-even scalar case discussed here is very different from the axion (or axion-like particle) case, in which the axion couplings to photons, electrons and nucleons are all independent parameters~\cite{Irastorza:2018dyq}, whereas for $S$, all the couplings are governed by the single parameter $\sin\theta$, thus making it more predictable.

\section{Luminosity limits}
\label{sec:limits}

\begin{table}[!t]
    \centering
    \caption{Stellar parameters for the Sun, RGs, {WDs} and HB stars, adopted in this paper: the dominant elements in the core and their mass fractions, the core temperatures $T$, the electron number density $n_e$, the size $R$ , and the luminosity limits in unit of the solar luminosity of ${\cal L}_\odot = 4 \times 10^{33}$ erg$\cdot$sec$^{-1}$.
    %For the Sun, the luminosity limit can be constrained down to 10\% of the observed luminosity, as shown in parenthesis.
    }
    \label{tab:stars}
        \begin{tabular}{l|c|c|c|c|c} \hline\hline
    %\multicolumn{4}{|c|}{Parameters/Masses} \\\hline
    Star & Core composition & $T$ [keV]  & $n_e \; [{\rm cm}^{-3}]$  & $R$ [cm] & ${\cal L}/{\cal L}_\odot$ \\ \hline\hline
    \multirow{2}{*}{Sun~\cite{Redondo:2013lna, Song:2017kvf, Guarini:2020hps}} & 75\% H  & \multirow{2}{*}{1} &
    \multirow{2}{*}{$10^{26}$} & \multirow{2}{*}{$7\times10^{10}$}  &
    \multirow{2}{*}{{0.03}} \\
    & 25\% $^4$He &&& \\ \hline
    RGs~\cite{Viaux:2013lha}  & $^{4}$He  & 10 & $3\times 10^{29}$  & $6\times 10^{8}$ &
    2.8 \\ \hline
    \multirow{2}{*}{WDs~\cite{Raffelt:1996wa, Blinnikov:1990ui, Panotopoulos:2020kuo, whitedwarf}}   & 50\% $^{12}$C   &
    \multirow{2}{*}{6}  & \multirow{2}{*}{$10^{30}$}   &
    \multirow{2}{*}{$10^{9}$}  & \multirow{2}{*}{$10^{-5}$ to 0.03} %~\cite{Panotopoulos:2020kuo,whitedwarf}}
    \\ %\hline
    & 50\% $^{16}$O &&& \\ \hline
    HB stars~\cite{Ayala:2014pea} & $^{4}$He  & 8.6 & $3\times 10^{27}$  & $3.6 \times 10^{9}$ &
    5 \\
    \hline\hline
    \end{tabular}
\end{table}

In Table~\ref{tab:stars}, we summarize the various stellar parameters, such as the elemental composition, core temperature $T$, electron number density $n_e$, the size $R$ and the observed luminosity limits, for the Sun, RGs, {WDs}, and HB stars adopted in this paper for the luminosity limit calculation. The Sun is the best-measured star and modern solar models constrain new forms of energy loss down to a few percent of the measured solar luminosity ${\cal L}_\odot = 4 \times 10^{33}$ erg$\cdot$sec$^{-1}$~\cite{Raffelt:1996wa, Redondo:2013lna, Gondolo:2008dd, Vinyoles:2015aba, Song:2017kvf}. We will use 3\% of the observed solar luminosity~\cite{Song:2017kvf, Guarini:2020hps} to derive the solar limits in the next section.

As for the RGs, just before helium ignition, their cores become hot and compact at the center of the huge stellar atmosphere. As the envelope temperature is much lower than in the RG core, we make the conservative assumption that the scalar $S$ can be effectively produced only in the core. The main energy loss is by neutrino emission at this stage of stellar evolution. New energy loss mechanisms are constrained to about $2.8{\cal L}_\odot$~\cite{Raffelt:1996wa, Viaux:2013lha}, as shown in Table~\ref{tab:stars}; higher energy losses would delay the onset of helium ignition, in disagreement with observations matching stellar models.

On the other hand, the HB star cores are puffed up due to energy released by fusion during the helium-burning stage, thus lowering the electron density. New energy loss mechanisms would cause the core to contract, heating it up and enhancing the rate of helium fusion, thus shortening the lifetime of the star. Comparing the measured helium-burning lifetime with standard stellar models, new energy losses are constrained to about $5{\cal L}_\odot$~\cite{Raffelt:1996wa, Ayala:2014pea}, as shown in Table~\ref{tab:stars}.

%Apart from these stellar objects, one can also consider very degenerate stellar remnants like WDs and neutron stars (NSs) to put constraints on the light scalar scenario~\cite{Raffelt:1996wa, Blinnikov:1990ui}. However, the emission of new particles in scattering processes in these electron-degenerate cores is highly suppressed by Pauli blocking. As a result, we expect the WD and NS limits to be weaker than the stellar limits derived in the next section, and hence, do not consider them in our analysis.

{Apart from these stellar objects, one can also consider very degenerate stellar remnants like WDs to put constraints on the light scalar scenario~\cite{Raffelt:1996wa, Blinnikov:1990ui}. In this case, one has to take into consideration the Pauli-blocking effect due to degenerate electron-cores, which can be captured by a multiplicative factor $F$ in the scalar emission rate. It turns out that for the environment of WDs the factor $F \simeq 1$~\cite{Nakagawa:1987pga, Nakagawa:1988rhp, Raffelt:1996wa}. For simplicity we will simply take $F = 1$ in our analysis. Furthermore, we assume the WD core to be 50\% $^{12}$C and 50\% $^{16}$O~\cite{Raffelt:1996wa, Blinnikov:1990ui}. Depending on the relevant parameters, the WD luminosities can span orders of magnitude. To be concrete, we adopt two luminosity limits, i.e. $10^{-5} {\cal L}_\odot$ and $0.3 {\cal L}_\odot$~\cite{Panotopoulos:2020kuo}, to set limits on the light scalar $S$.
}

While calculating the luminosity limit from Eq.~(\ref{eqn:lumi}), we have to take into consideration both the decay and reabsorption of $S$ inside the stellar core. This is especially relevant for larger values of $\sin\theta$, which may not allow the $S$ particles to escape the stellar core, thus rendering the luminosity limits inapplicable. For the light scalar with mass $m_S < 2m_e$, $S$ dominantly decays into two photons, %\footnote{The scalar $S$  could decay in principle also into two neutrinos via the $Z-Z-\nu$ loops, i.e. $S\to \nu\bar\nu$. However, following the calculations in Ref.~\cite{Goodsell:2017pdq}, this channel is highly suppressed by the tiny neutrino masses, and thus can be safely neglected.}
with the width given by Eq.~(\ref{eqn:width}). In our case, the decay of $S$ is relevant only when the mass $m_S \gtrsim 100 $ keV and $\sin\theta \sim 1$. For smaller masses and mixing angles, the reabsorption effect plays a crucial role.

After being produced, $S$ can be reabsorbed in the stellar core via the following processes:
\begin{itemize}
    \item Inverse bremsstrahlung process $e+ N + S \to e+N$.
    \item Inverse Compton process $e + S \to e + \gamma$.
    \item Inverse Primakoff process $X + S \to \gamma + X$ (with $X$ being either electron or nuclei).
\end{itemize}
{All these absorption processes above would be modified to some extent by the plasma effect~\cite{Gelmini:2020xir}. However, as in the production case (cf.~Fig.~\ref{fig:channels}), the plasma effect is expected to be subdominant for the generic scalar case, and the absorption of $S$ inside the stars} is dominated by the inverse bremsstrahlung process, due to the coupling $y_N$ being much larger than the electron Yukawa coupling $y_e$ and the loop-level coupling of $S$ to photons.

The corresponding inverse MFP can be calculated as follows~\cite{Burrows:1990pk, Giannotti:2005tn}:
\begin{eqnarray}
\label{eqn:mfp}
\lambda^{-1}_{} %& %\ \equiv \ &
%\frac{1}{2 E_S} \frac{d {\cal N}_S (-k_S)}{d \Pi_S} \nonumber \\
& \ = \ &
\frac{1}{2 E_S} \sum_i
\int {\rm d} \Pi_4 \sum_{\rm spins} |{\cal M}^\prime_i|^2 (2\pi)^4
\delta^4 (p_1 + p_2 - p_3 - p_4 + k_S) f_1^{(e)} f_2^{(N_i)} \,,
\end{eqnarray}
%with $p_{1,\, 2}$ the momenta of $e$ and $N_i$ in the initial state, $p_{3,\,4}$ the momenta of $e$ and $N_i$ in the final state, $k_S$ the momentum of $S$,
with ${\cal M}^\prime_i$ being the scattering amplitude for the $3\to2$ processes $e + N_i+ S \to e + N_i$ and ${\rm d} \Pi_4$ is the four-body phase space for the initial and final state nuclei $N_i$.
%and
%\begin{eqnarray}
%f^{(e,\,N_i)} = \frac{n_{e,\,N_i}}{2}
%\left( \frac{2\pi}{m_{e,\,N_i} T} \right)^{3/2}
%\exp \left\{ - \frac{{\bf P}_{e,\,N_i}}{m_{e,\, N_i} T}\right\} \,,
%\end{eqnarray}
%the Maxwell-Boltzmann distributions for 4lectron and nuclei in the initial state, with ${\bf P}_{e,\, N_i}$  the 3-momentum of electron and nuclei.
Following the calculations in Ref.~\cite{Dev:2020eam}, the inverse MFP can be simplified as
%\begin{eqnarray}
%\label{eqn:MFP2}
%\lambda^{-1} = \sum_N \frac{2 \pi^{1/2} Z^2 A^2 \alpha^2 y_{N}^2 \sin^2\theta n_e n_N}{ m_e^{3/2} T^{7/2}x^2}
%\int du dv dz \sqrt{uv} e^{-u} \frac{\delta (u-v+x)}{(u+v-2\sqrt{uv}z)^2} \,,
%\end{eqnarray}
\begin{eqnarray}
\label{eqn:MFP2}
\lambda^{-1}_{} & = & \left( \sum_i Z_{N_i}^2 A_{N_i}^2 n_{N_i} \right) \frac{2 \pi^{1/2}  \alpha^2 y_{N}^2 \sin^2\theta n_e}{ m_e^{3/2} T^{7/2}x^2}
\int_{0}^{\infty} du \int_q^\infty dv \int_{-1}^1 dz \nonumber \\
&& \qquad \times
\sqrt{uv} e^{-u} \frac{\delta (u-v+x)}{(u+v-2\sqrt{uv}z)^2} \,,
\end{eqnarray}
where we have summed over all the nuclei for coherent scattering.\footnote{The temperature dependence of the emission rate in Eq.~\eqref{eqn:QB:eN} and the MFP in Eq.~\eqref{eqn:MFP2} are different from those in the supernova case~\cite{Dev:2020eam} because the production (reabsorption) in the stellar core is dominated by the (inverse) $e-N$ bremsstrahlung mediated by a photon, whereas in the supernova core, it is dominated by the (inverse) $N-N$ bremsstrahlung mediated by a pion, with different Lorentz structures in the amplitude.} As the MFP $\lambda$ is a function of the scalar energy $E_S$, we average over the distribution of $S$ to obtain an effective energy-independent MFP~\cite{Ishizuka:1989ts} that goes into Eq.~\eqref{eqn:Pabs}:
\begin{eqnarray}
\langle \lambda_{}^{-1} \rangle \ \equiv \
\frac{\bigintss {\rm d} E_S \frac{E_S^3}{e^{E_S/T}-1} \lambda^{-1}_{} (E_S)}{\bigintss {\rm d} E_S \frac{E_S^3}{e^{E_S/T}-1}}
\ = \ \frac{\bigintss {\rm d} x \frac{x^3}{e^{x}-1} \lambda_{}^{-1} (x)}{\bigintss {\rm d} x \frac{x^3}{e^{x}-1}} \,.
\label{eqn:MFP3}
\end{eqnarray}
%Then we can insert the factor of
%\begin{eqnarray}
%P_{\rm abs} = \exp \left\{ -  \frac{R}{ \langle \lambda_{\rm B} \rangle } \right\}
%\end{eqnarray}
%into Eq.~(\ref{eqn:QB:eN}) to account for the absorption of $S$ inside the stars.
\begin{figure*}[t!]
    \centering
    \includegraphics[width=0.55\textwidth]{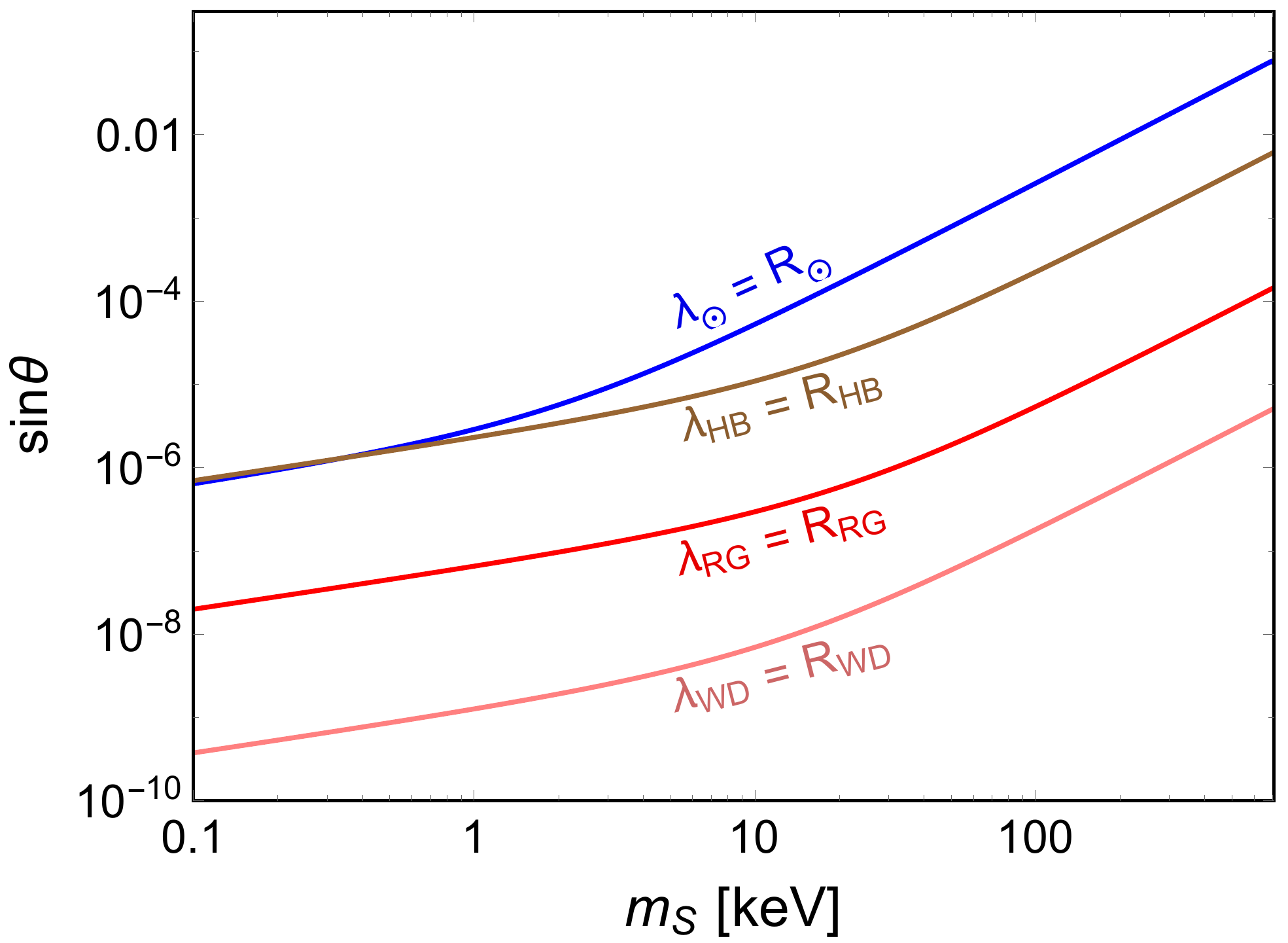}
    \caption{Contours of the scalar MFP $\lambda_\odot = R_\odot$ in the Sun (blue), $\lambda_{\rm RG} = R_{\rm RG}$ in RGs (red), {$\lambda_{\rm WD} = R_{\rm WD}$ in the WDs (pink)}, and $\lambda_{\rm HB} = R_{\rm HB}$ in HB stars (brown),  as a function of the scalar mass $m_S$ and mixing angle $\sin\theta$.}
    \label{fig:mfp}
\end{figure*}

The contours of MFP values equal to the sizes of the stellar cores, i.e. $\lambda_\odot = R_\odot$ in the Sun,  $\lambda_{\rm RG} = R_{\rm RG}$ in the RGs, {$\lambda_{\rm WD} = R_{\rm WD}$ in the WDs,} and $\lambda_{\rm HB} = R_{\rm HB}$ in the HB stars (cf.~Table~\ref{tab:stars}) are shown in Fig.~\ref{fig:mfp} respectively as the blue, red, pink and brown lines, as a function of the scalar mass $m_S$ and mixing angle $\sin\theta$.  For other values of MFP, we need only to rescale the mixing angles by $\sin^{2}\theta$. We can apply the energy loss arguments discussed in the previous section to set limits on the scalar mass and mixing only when the MFP exceeds the stellar (core) size; for smaller values of the MFP, the produced $S$ particles will be trapped inside the core and the corresponding luminosity will be exponentially suppressed (cf.~Eq.~\eqref{eqn:Pabs}).

\begin{figure*}[t!]
    \centering
   % \begin{tabular}{lr}
    \includegraphics[width=0.55\textwidth]{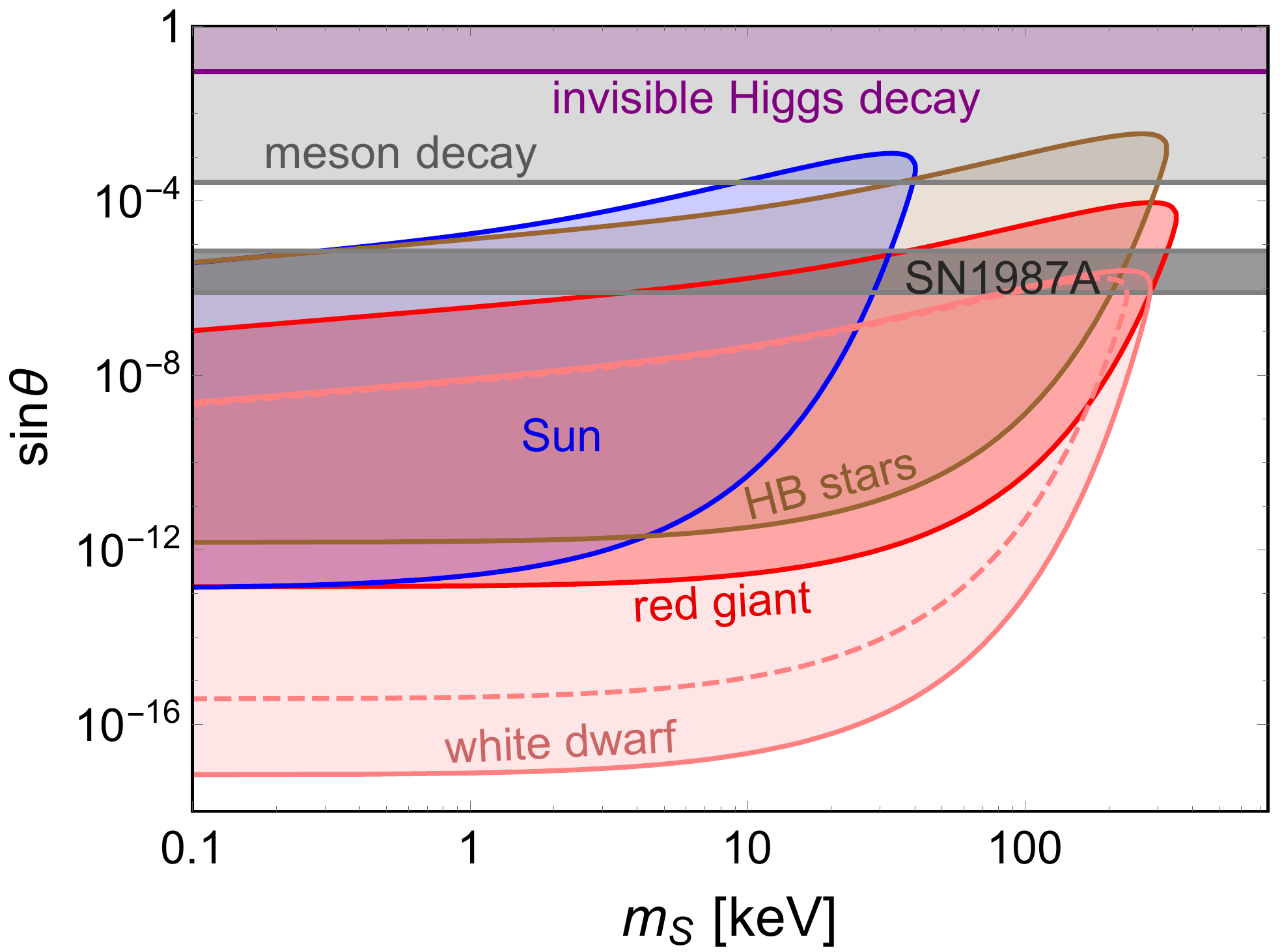}
    %\includegraphics[width=0.5\textwidth]{scan.pdf}
     % \end{tabular}
    \caption{Stellar luminosity limits on the scalar mass $m_S$ and mixing angle $\sin\theta$ from the Sun (blue), RGs (red),  {WDs (pink)} and HB stars (brown), with the stellar parameters shown in Table~\ref{tab:stars}.  {The solid and dashed pink curves correspond respectively to the WD luminosity limits of $10^{-5} {\cal L}_\odot$ and $0.03{\cal L}_\odot$.}
    Also shown in this figure are the SN1987A limit with the luminosity of $3\times10^{53}$ erg$\cdot$sec$^{-1}$ (dark gray)~\cite{Dev:2020eam}, invisible meson decays (light gray)~\cite{Dev:2017dui,Dev:2019hho} and the current LHC limits on invisible decay of the SM Higgs (purple)~\cite{Sirunyan:2018owy, ATLAS:2020cjb}.  }
    \label{fig:limits}
\end{figure*}

With the decay and reabsorption of $S$ taken into consideration, the luminosity limits of the Sun, RGs, WDs and HB stars on the scalar mass $S$ and mixing angle $\sin\theta$ are presented in Fig.~\ref{fig:limits} respectively as the blue, red, pink and brown shaded regions.
%For the solar limits, the solid and dashed blue lines correspond respectively to the conservative and aggressive luminosity limits of ${\cal L}_\odot$ and $0.1 {\cal L}_\odot$ (cf.~Table~\ref{tab:stars}).
%The observed WD luminosities span a broad range, from $10^{-5} {\cal L}_\odot$ to $0.03 {\cal L}_\odot$~\cite{Panotopoulos:2020kuo,whitedwarf}, which are denoted respectively by the solid and dashed pink curves in Fig.~\ref{fig:limits}.
The resulting excluded $m_S$ and $\sin\theta$ ranges from the stars are collected in Table~\ref{tab:limits}.

\begin{table}[!t]
    \centering
    \caption{Luminosity limits of the Sun, RGs, {WDs} and HB stars on a generic light scalar $S$, with the third column showing the excluded $\sin\theta$ ranges, and the fourth column showing the excluded ranges for $m_S$. The other stellar parameters are taken from Table~\ref{tab:stars}. See text and Fig.~\ref{fig:limits} for more details.}
    \label{tab:limits}
       \begin{tabular}{c|c|c|c} \hline\hline
    %\multicolumn{4}{|c|}{Parameters/Masses} \\\hline
    Star & Luminosity limit [${\cal L}_\odot$] & $\sin\theta$ range & $m_S$ range \\ \hline \hline
    %\multirow{2}{*}{Sun} & 1 & $7.9\times10^{-13} - 9.8\times10^{-4}$   &
    %$<36.4$ keV \\
    %& 0.1 & $2.5\times10^{-13} - 1.1\times10^{-3}$  &
    %$<38.8$ keV  \\ \hline
    Sun & 0.03 & {$1.4\times10^{-13} - 1.2\times10^{-3}$}  &
    {$<40.2$ keV}  \\ \hline
    RGs & 2.8 & $1.4 \times10^{-13} - 9.0 \times10^{-5}$ & $<350$ keV \\ \hline
    \multirow{2}{*}{WDs} & 0.03 & $3.8 \times10^{-16} - 1.7\times10^{-6}$ & $<232$ keV \\
    & $10^{-5}$ & $7.0 \times10^{-18} - 2.5\times10^{-6}$ & $< 282$ keV \\ \hline
    HB stars & 5 &  $1.5 \times10^{-12} - 3.4\times10^{-3}$ & $<323$ keV \\
    \hline\hline
    \end{tabular}
\end{table}

For comparison, we also show the following laboratory and astrophysical limits in Fig.~\ref{fig:limits}:
\begin{itemize}
    \item {\it Supernova}: In the supernova core, the light scalar can be produced from the nucleon-nucleon bremsstrahlung process $N + N \to N + N + S$ with $N = p,\, n$ and the scalar $S$ couples either to the nucleons or the pion mediator. Our recent calculations in Ref.~\cite{Dev:2020eam} has revealed an unusual cancellation of the production diagrams of $S$, which is very different from the axion or dark photon case. In the limit of $m_S \to 0$, the SN1987A luminosity limit of $3\times10^{53}$ erg$\cdot$sec$^{-1}$ excludes the darker gray shaded region in Fig.~\ref{fig:limits}, with $7.8 \times 10^{-7} \lesssim \sin\theta \lesssim 7.0 \times10^{-6}$.

    \item {\it Meson decay}: Through mixing with the SM Higgs, the scalar $S$ has loop-level flavor-changing neutral current (FCNC) couplings to the SM quarks and therefore can be produced from the FCNC decays of the SM mesons, such as $K \to \pi + S$. The current limits from NA48/2~\cite{Batley:2009aa,Batley:2011zz}, E949~\cite{Artamonov:2009sz}, KOTO~\cite{Ahn:2018mvc}, NA62~\cite{Ceccucci:2014oza,Ruggiero:2019}, KTeV~\cite{AlaviHarati:2003mr,AlaviHarati:2000hs,Alexopoulos:2004sx, Abouzaid:2008xm}, BaBar~\cite{Aubert:2003cm,Lees:2013kla}, Belle~\cite{Wei:2009zv}, and LHCb~\cite{Aaij:2012vr} require that the mixing angle $\sin\theta <2.6 \times 10^{-4}$, which is shown as the light gray shaded region in Fig.~\ref{fig:limits}. See Refs.~\cite{Dev:2017dui,Dev:2019hho} for more details.

    \item  {\it Invisible Higgs decay}: The current precision Higgs measurements at the LHC set an upper bound on the invisible branching fraction of the SM, i.e. ${\rm BR} (h \to {\rm inv.}) < 0.19$~\cite{Sirunyan:2018owy, ATLAS:2020cjb}.  From the trilinear scalar coupling $\lambda_{hhh} = 3 m_h^2/v_{\rm EW}$ in the SM (with $m_h = 125$ GeV for the SM Higgs mass and $v_{\rm EW} \simeq 246$ GeV  for the electroweak VEV) and  the $h-S$ mixing, we can obtain the partial width for the decay (in the limit $m_h\gg m_S$)
    \begin{eqnarray}
    \Gamma (h \to SS) = \frac{9 m_h^3 \sin^4\theta}{8 \pi v_{\rm EW}^2} \, .
    \end{eqnarray}
    This implies that, to satisfy the invisible Higgs decay limit, the mixing angle $\sin\theta < 0.09$, which is denoted by the purple shaded region in Fig.~\ref{fig:limits}.
\end{itemize}

As shown in Fig.~\ref{fig:limits} and Table~\ref{tab:limits}, the luminosity limits from the Sun, RGs, WDs and HB stars exclude a broad range of parameter space for the generic light CP-even scalar $S$, ranging from roughly {$10^{-18}$} to $10^{-3}$ for the mixing angle $\sin\theta$ and up to $\sim 350$ keV for the scalar mass $m_S$. The stellar limits are largely complementary to those from supernovae, meson decays and the precision Higgs data.

We should note here that the scalar $S$ produced in the Sun can also be absorbed by the atoms in the detectors set up for DM direct detection, such as XENON1T and LUX (see more details in Section~\ref{sec:excess}). The electron recoils will generate prompt scintillation light and ionization events in the detectors, i.e. the so-called ``S1'' and ``S2'' signals in the Xe detectors. The recent XENON1T~\cite{Aprile:2018dbl,Aprile:2019xxb} and LUX~\cite{Akerib:2017uem} data can then used to set limits on the coupling of $S$ to electron, or effectively on the mixing angle $\sin\theta$ with the SM Higgs~\cite{Budnik:2019olh}. However, the limits presented in Fig.~3 of Ref.~\cite{Budnik:2019olh} seem to be incomplete, i.e. there should not only be lower bounds on the coupling of $S$ to electron, but also upper bounds. Analogous to the stellar limits shown in Fig.~\ref{fig:limits}, for sufficiently large coupling, the scalar $S$ will be trapped in the Sun and thus can not reach the detectors on the Earth. As the DM detector simulations and the resultant Xe detector limits on $S$ are beyond the main scope of this paper, we will not include the XENON1T and LUX limits in Fig.~\ref{fig:limits} and in the following.

%that in the low-mass range with $m_S \lesssim 200$ keV, the stellar limits from the Sun, WDs, RGs and HB stars are much more stringent than that from SN1987A, meson decays and the precision Higgs data.

%\blue{limits of Sun, Red Giants and white dwarfs on $m_S$ and $\sin\theta$, and fitting XENON1T excess....}

A light $S$ might also contribute to the relativistic degrees of freedom $N_{\rm eff}$ in the early universe, thus be constrained by the current precision Planck data~\cite{Aghanim:2018eyx}. In addition, if the scalar lifetime $\tau_S \gtrsim 1$ sec, and it remains in equilibrium with the SM particles at the big bang nucleosynthesis (BBN) epoch, the scalar $S$ will affect the primordial abundance of light elements~\cite{Kainulainen:2015sva, Fradette:2017sdd}. For the parameter space we are interested in, i.e. $\sin\theta \lesssim 0.1$ and $m_S \lesssim$ 100 keV,  we find that the scalar $S$ never comes into equilibrium with the SM sector, and moreover, the decay rate $\Gamma (S \to \gamma\gamma)$ is significantly smaller than the Hubble expansion rate ${\cal H} \simeq 10 T_{\rm BBN}^2/M_{\rm Pl}$, with $T_{\rm BBN} \sim $ MeV the BBN temperature and  $M_{\rm Pl}$ the Planck mass. Therefore we do not have any cosmological limits on $S$ in Fig.~\ref{fig:limits}.

\section{XENON1T excess}
\label{sec:excess}
In this section, we examine the possibility if a light CP-even scalar $S$ emitted from the Sun can explain the keV-scale excess in the electron recoil events recently observed in the XENON1T experiment~\cite{Aprile:2020tmw}. To  this end, we need the differential production rate $N_S$ for $S$ in the Sun with respect to its energy $E_S$, which can be calculated from the $e-N$ bremsstrahlung production rate given in Eq.~\eqref{eqn:rate:master}, multiplied by the decay and absorption probabilities (cf.~Eqs.~\eqref{eqn:Pdecay} and \eqref{eqn:Pabs}):
\begin{eqnarray}
\frac{dN_S}{d E_S} = \sum_i \frac{d}{dE_S}
\int {\rm d} \Pi_5 \sum_{\rm spins} |{\cal M}_i|^2 (2\pi)^4
\delta^4 (p_1 + p_2 - p_3 - p_4 - k_S) f_1^{(e)} f_2^{(N_i)} P_{\rm decay} P_{\rm abs} \, .
\end{eqnarray}
The calculation is very similar to Eq.~(\ref{eqn:QB:eN}), and we get
\begin{eqnarray}
\frac{dN_S}{d E_S} &=&
\left(\sum_i  Z_{N_i}^2 A_{N_i}^2 n_{N_i} \right) \frac{ \alpha^2 y_{N}^2 \sin^2\theta n_e }{\pi^{3/2} m_e^{3/2} T^{3/2}} \nonumber \\
&& \qquad \times \int_{q}^{\infty} du \int_0^\infty dv \int_{-1}^1 dz \sqrt{uv} e^{-u}  \sqrt{1 - \frac{q^2}{x^2}} \frac{\delta (u-v-x)}{(u+v-2\sqrt{uv}z)^2} P_{\rm decay} P_{\rm abs}\,.
\end{eqnarray}
Then the differential number density of $S$ at the Earth is
\begin{eqnarray}
\label{eqn:dnSdES}
\frac{dn_S}{dE_S} = \frac{V_\odot}{4\pi D^2} \frac{ dN_S}{d E_S} \,,
\end{eqnarray}
with $V_\odot=\frac{4\pi}{3}R_\odot^3$ being the solar volume and $D=1~{\rm AU}=1.5\times10^{13}$ cm the distance from the Sun to the Earth.
%and $V_\odot = \frac{4\pi}{3} R_\odot^3$ the volume of Sun.

Using Eq.~(\ref{eqn:width}), we find that the decay length of $S$ is much longer than the Sun-Earth distance for the parameter space of XENON1T excess, and therefore, the decay of $S$ can be neglected. Similarly, the absorption of $S$ inside the Earth can be neglected due to the following reason: the MFP of $S$ inside the Earth can be written as
\begin{eqnarray}
\lambda_\oplus^{-1} %\simeq \sum_i n_i \sigma_{Se}^{i}
\simeq n_{e,\, \oplus} \sigma_{Se}^{} \,,
\end{eqnarray}
where $n_{e,\,\oplus} \sim m_\oplus/2m_p V_\oplus$ the electron number density in the Earth, with $m_\oplus$ the Earth mass, $V_\oplus$ the Earth volume, $m_p$ the proton mass, and the cross section for the absorption~\cite{Avignone:1986vm}
\begin{eqnarray}
\label{eqn:sigma_Se}
\sigma_{Se}^{} \simeq \frac{y_e^2 \sin^2\theta}{4\pi\alpha} \sigma_{\gamma e}^{} \,,
\end{eqnarray}
with $\sigma_{\gamma e} \sim$ Mb the photoelectric cross section~\cite{Veigele:1973tza}. Then the MFP of $S$ inside the Earth can be estimated as
\begin{eqnarray}
\lambda_\oplus \simeq 6 \times 10^{17} \, {\rm km} \;
\left(\frac{\sin\theta}{10^{-11}}\right)^{-2}
\left(\frac{\sigma_{\gamma e}}{\rm Mb}\right)^{-1} \,,
\end{eqnarray}
which is much larger than the Earth size.
%, because the MFP

\begin{figure}[t!]
    \centering
    \includegraphics[height=0.365\textwidth]{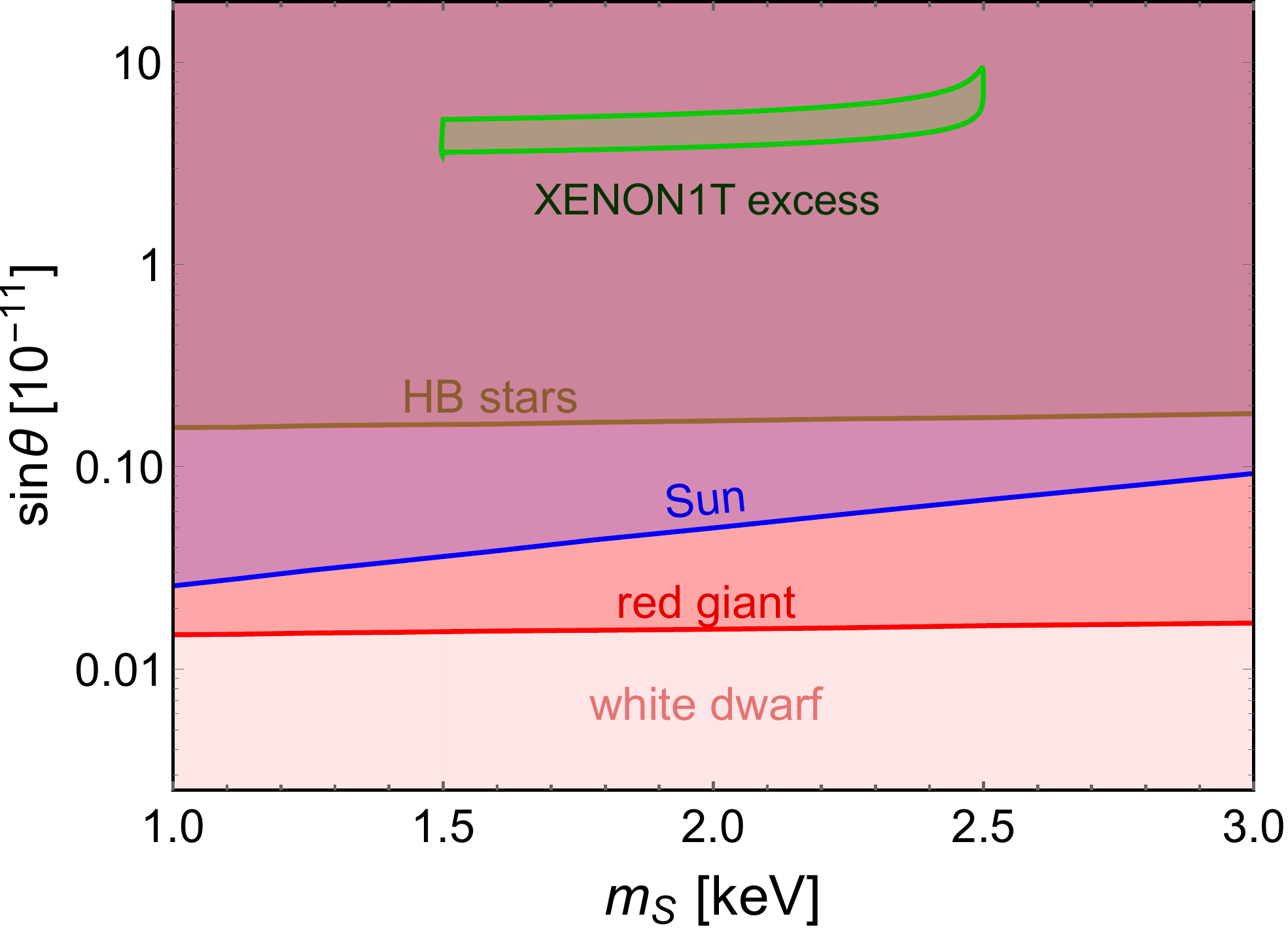}
    \caption{Parameter space of $m_S$ and $\sin\theta$ at the $2\sigma$ C.L. for the XENON1T excess (shaded green). The stellar exclusion limits from Fig.~\ref{fig:limits} are also shown for comparison. }
    \label{fig:excess}
\end{figure}

When $S$ passes through the Earth to reach the underground XENON1T detector, it may be absorbed by electrons in the Earth, which is very similar to the axioelectric process~\cite{Avignone:1986vm}, and can thereby generate the anomalous keV-scale electron recoil events.
%This is induced by the coupling of $S$ to electrons, although the couplings to nucleus are much larger.
%via the process $e + N_i + S \to e + N_i$, with $N_i$ the nucleus involved.
%but it is dominated by the coupling of $S$ to nucleus, which is much larger than the couplings of $S$ to electrons.
%In the XENON1T detector, the scalar $S$ can be absorbed by electrons in the XENON detector, which is similar to what happened when $S$ passes through the Earth,
%the process $e + {\rm Xe} + S \to e + {\rm Xe}$, with Xe standing for the XENON nuclei.
%As for the absorption of $S$ inside Earth, this process is dominated by the coupling of $S$ to nucleus,
%and the cross section is given by Eq.~(\ref{eqn:sigma_Se}), with $\sigma_{Se}$ by the photoelectric cross section $\sigma_{Se}^{\rm (Xe)}$ for XENON atom.
We can readily estimate the differential event rate at the XENON1T detector:
\begin{eqnarray}
\label{eqn:dRdER}
\frac{dn_R}{dE_R} = \epsilon_{\rm det} n_{\rm Xe} \sigma_{Se}^{\rm (Xe)} \frac{dn_S}{dE_S} \,,
\end{eqnarray}
with $E_R$ being the recoil energy, $\epsilon_{\rm det}$ the detector efficiency~\cite{Aprile:2020tmw}, $n_{\rm Xe} = 4.2 \times 10^{27}/$ton, and $\sigma_{Se}^{\rm (Xe)}$ the cross section in Eq.~(\ref{eqn:sigma_Se}) for the xenon element in the detector. The XENON1T excess parameter space of $m_S$ and $\sin\theta$ at the $2\sigma$ confidence level (C.L.) is shown in Fig.~\ref{fig:excess} as the green shaded region. As shown in this figure, the XENON1T excess requires that  $m_S \sim $ keV and $\sin\theta \sim 10^{-11}$, which however has been excluded by the stringent luminosity limits from the Sun, RGs, WDs and HB stars. Thus we conclude that the light CP-even scalar cannot explain the XENON1T excess, similar to the CP-odd scalar (axion) case.
%The XENON1T excess region is also indicated by  the green shaded area in Fig.~\ref{fig:limits}. As seen clearly in Fig.~\ref{fig:limits}, the XENON1T excess region has been excluded by the limits from White dwarfs.

%The XENON1T preferred $2\sigma$ region is shown as the green shaded region in Fig.~\ref{fig:limits}, with which however has been excluded by the stringent stellar limits.

% with the process $e + N + S \to e + N$, with the S coupling predominately to nuclei $N$.
%with the cross section~\cite{Avignone:1986vm}
%\begin{eqnarray}
%\sigma_{Se}^{\rm (Xe)} = \frac{A_{\rm Xe}^2 y_N^2 \sin^2\theta}{4\pi\alpha} %\sigma_{\gamma e}^{\rm (Xe)} \,,
%\end{eqnarray}

%\section{Simple realistic models}

\section{Real singlet scalar model}
\label{sec:singlet}
\begin{figure*}[t!]
    \centering
    \includegraphics[width=0.6\textwidth]{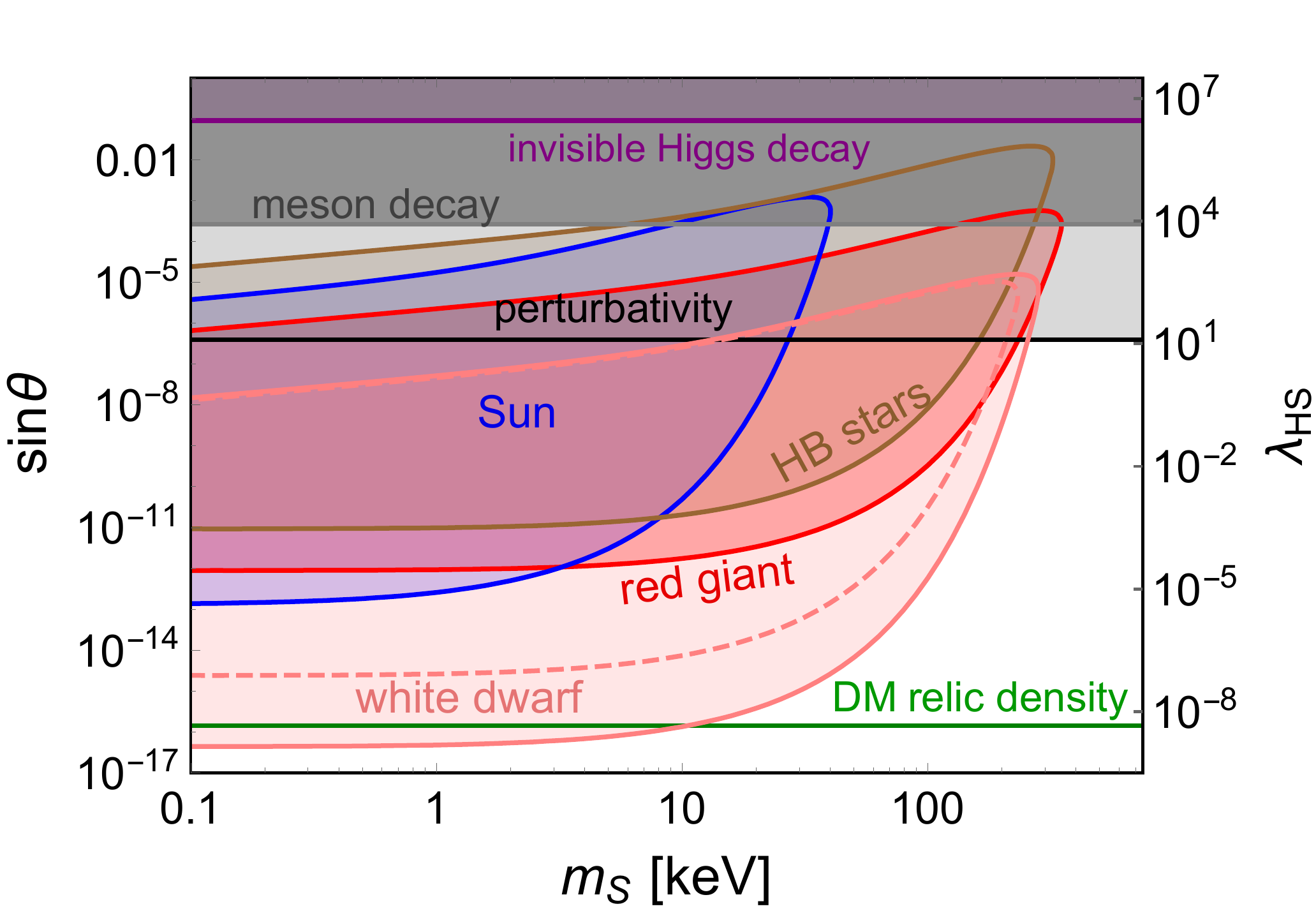}
    \caption{The same as in Fig.~\ref{fig:limits}, but for the scalar $S$ in the real singlet model, as function of $m_S$ and $\sin\theta$ (or effectively the quartic coupling $\lambda_{HS}$), with the singlet VEV $v_S = 2$ keV. For this model we do not have the SN1987A limit (see text for details). The horizontal dark green line corresponds to the observed DM relic density of $\Omega_{S}h^2=0.12$ [see Eq.~(\ref{eqn:DM})]. In the shaded region above the horizontal black line the quartic coupling $\lambda_{HS} > 4\pi$.  }
    \label{fig:limits:singlet}
\end{figure*}

In this section, we consider a simple realistic model for the light scalar $S$, i.e. extending the SM by adding a real singlet scalar $S$. The most general scalar potential  invariant under a $Z_2$ symmetry under which $S\to -S$ can be written as
\begin{eqnarray}
\label{eqn:potential}
{\cal V} = \frac{\lambda_H}{4} \left( H^\dagger H - v_{\rm EW}^2 \right)^2 +
\frac{\lambda_S}{4} \left( S^2 - v_S^2 \right)^2 +
\frac{\lambda_{HS}}{2} \left( H^\dagger H - v_{\rm EW}^2 \right) \left( S^2 - v_{S}^2 \right) \,,
\end{eqnarray}
with $v_S$ being the VEV of $S$.
If the $S$ mass is at the keV-scale, it could be a DM candidate, and as shown in Ref.~\cite{Babu:2014pxa}, could explain the anomalous X-ray spectrum at 3.55 keV~\cite{Bulbul:2014sua, Boyarsky:2014jta}. In the early universe, the scalar $S$ can be produced either from the decay of the SM Higgs $h \to SS$, or from scattering of two SM particles $hh,\, WW,\, ZZ,\, t\bar{t} \to SS$. The subsequent annihilation and decay of $S$ can never reach equilibrium with the SM plasma, which makes the light scalar $S$ a natural freeze-in DM~\cite{Hall:2009bx}. All the channels are through the SM Higgs portal, and the relic density of $S$ is related to the quartic coupling $\lambda_{HS}$ via~\cite{Babu:2014pxa}
\begin{eqnarray}
\label{eqn:DM}
\Omega_{S} h^2 \simeq 0.12 \left( \frac{\lambda_{HS}}{4.5\times10^{-9}} \right)^{2} \,.
\end{eqnarray}
The observed DM relic density $\Omega_{\rm DM}^{\rm obs} h^2 = 0.120\pm 0.001$~\cite{Aghanim:2018eyx} is shown as the dark green line in Fig.~\ref{fig:limits:singlet}.

%\begin{figure*}[t!]
%    \centering
%    \includegraphics[width=0.4\textwidth]{diagram.pdf}
%    \caption{Feynman diagram for the production of $S$ in the stars with $T > u$, with the $S$ coupling to the either $(a)$ or $(b)$ of the nucleus $N$.}
%    \label{fig:diagram}
%\end{figure*}

As the stellar temperatures are all at the keV-scale, the VEV $v_S$ of the scalar $S$ might be below or above the stellar temperatures $T$. If $v_S > T$, the mixing of $S$ with the SM Higgs is related to the parameters in the potential (\ref{eqn:potential}) via
\begin{eqnarray}
\label{eqn:sintheta}
\sin\theta \simeq \frac{\sqrt2 \lambda_{HS} v_{\rm EW} v_S}{m_h^2} \,.
\end{eqnarray}
If the stellar temperatures $T>v_S$, the $S$ field does not develop a non-vanishing VEV inside the star, and the scalar $S$ couples to the SM Higgs through the $\lambda_{HS}$ term but does not mix with it. As a result, $S$ can not couple directly to the SM fermions. However, it can be pair produced from the $2\to4$ bremsstrahlung process
\begin{eqnarray}
\label{eqn:2to4}
e + N \to e + N + S + S \,,
\end{eqnarray}
which is mediated by the trilinear coupling of SM Higgs $h$ to $S$. Compared to the $2\to3$ process in Eq.~(\ref{eqn:brem2}), the phase space in this $2\to4$ process is suppressed by a factor of $4\pi^2$, and the production amplitude square is roughly rescaled by a factor of
\begin{eqnarray}
\label{eqn:ratio}
\frac{\left( \sqrt2 \lambda_{HS} v_{\rm EW} \right)^2 {\bf p}_{\bf S}^2}{m_h^4 \sin^2\theta}
= \frac{{\bf p}_{\bf S}^2}{v_S^2} \,,
\end{eqnarray}
with the factor of $\sqrt2 \lambda_{HS}v_{\rm EW}$ from the $h-S-S$ vertex, $1/m_h^4$ from the SM Higgs propagator, ${\bf p_S}$ the 3-momentum of $S$, and we have used the relation in Eq.~(\ref{eqn:sintheta}). As the ratio ${\bf p}_{\rm S}^2/v_S^2$ in Eq.~(\ref{eqn:ratio}) is expected to be of order one, the energy loss due to the pair production of $S$ in Eq.~(\ref{eqn:2to4}) is less than the standard single production of $S$ in  Eq.~(\ref{eqn:brem2}) by roughly a factor of $4\pi^2$.

To fit the XENON1T excess~\cite{Aprile:2020tmw}, we can set the parameters to be
\begin{eqnarray}
\sin\theta \simeq 10^{-12.5}
\left(\frac{\lambda_{HS}}{6\times10^{-4}}\right)
\left(\frac{v_S}{1\, {\rm keV}}\right) \,.
\end{eqnarray}
Depending on the specific value of the singlet VEV $v_S$, the stellar limits and the XENON1T excess parameter space could differ to some extent. Here we consider three distinct scenarios:
\begin{itemize}

    \item
    If the singlet VEV $v_S$ lies above all the stellar temperatures, i.e. $v_S \gtrsim 10$ keV, we can relate the effective $h-S$ mixing angle $\sin\theta$ to the parameters in the potential (\ref{eqn:potential}) via Eq.~(\ref{eqn:sintheta}), and the stellar luminosity limits and the XENON1T excess parameter space will be the same as shown in Figs.~\ref{fig:limits} and \ref{fig:excess} respectively. Given Eq.~(\ref{eqn:sintheta}), one can also transfer the limits on the mixing angle $\sin\theta$ to the limits on the quartic coupling $\lambda_{HS}$ for a fixed value of singlet VEV $v_S$, which is shown in Fig.~\ref{fig:limits:singlet}.

    \item
    If $v_S$ is below all the stellar temperatures, i.e. $v_S \lesssim 1$ keV, as just aforementioned, the single production of $S$ via the bremsstrahlung process in Eq.~(\ref{eqn:brem2}) will be forbidden, and $S$ will be predominately pair-produced via the channel in Eq.~(\ref{eqn:2to4}). As a result, both the stellar limits in Fig.~\ref{fig:limits} and the XENON1T excess region in Fig.~\ref{fig:excess} will scale up universally by roughly a factor of $\sqrt{4\pi^2} = 2\pi$.

    \item
    The singlet VEV $v_S$ could also lie in between the stellar temperatures, of which the most phenomenologically interesting one is the case with the VEV $v_S$ above the solar temperature but below all other stellar temperatures, i.e. $1\, {\rm keV} \lesssim v_S \lesssim 10$ keV. Then the solar limits and the XENON1T excess region are the same as shown respectively in Figs.~\ref{fig:limits} and \ref{fig:excess}, whereas all other stellar limits are weakened by a factor of $2\pi$ compared to those in Fig.~\ref{fig:limits}. As an explicit example, the benchmark scenario with $v_S = 2$ keV is presented in Fig.~\ref{fig:limits:singlet}. Based on Eq.~(\ref{eqn:sintheta}), when the mixing angle is larger than $3.9 \times10^{-7}$, the quartic coupling $\lambda_{HS}$ becomes non-perturbative, i.e. $\lambda_{HS} > 4\pi$,. This is indicated by the horizontal black line in Fig.~\ref{fig:limits:singlet}.
    The temperature in the supernova core is $T_{\rm SN} = 30$ MeV, much larger than the VEV $v_S$; then the production of $S$ in supernova cores will be dominated by the pair production process $N + N \to N + N + S + S$, with $N$ here denoting nucleons. As for the case of Eq.~(\ref{eqn:2to4}) above, the production rate of $S$ will be roughly a factor of $4\pi^2$ times smaller than the standard bremsstrahlung process $N + N \to N + N + S$ in the supernova cores. Following the calculation procedure in Ref.~\cite{Dev:2020eam}, the SN1987A limits can not exclude any parameter space of $\sin\theta$ in the limit of $m_S \to 0$ for $v_S \sim {\rm keV}$. The meson decay and invisible Higgs decay limits are at the temperature of $T= 0$ and they are the same as in Fig.~\ref{fig:limits}. {As can be seen from Fig.~\ref{fig:limits:singlet}, the parameter space explaining the DM relic density is constrained by the most stringent WD limit, which imposes a lower limit of $m_S\gtrsim 10$ keV.}
\end{itemize}

%However, for sufficiently small $\sin\theta$ the light scalar $S$ cannot be kept in equilibrium with the SM photon.
%In particular, at the big bang nucleosynthesis (BBN) epoch with temperature $T_{\rm BBN} \sim {\rm MeV}$, if the mixing angle $\sin\theta \lesssim 0.1$ and the scalar mass $m_S \ll$ 100 keV, the decay rate $\Gamma (S \to \gamma\gamma) \exp [ -m_{S}/T_{\rm BBN} ]$ is significantly smaller than the Hubble expansion rate ${\cal H} \simeq 10 T_{\rm BBN}^2/M_{\rm Pl}$, with $M_{\rm Pl}$ the Planck mass.
%with $T_\ast \sim T_{\rm BBN} \sim $ MeV and
%The yellow shaded region in Fig.~\ref{fig:limits:singlet} is excluded by the condition that
%\begin{eqnarray}
%\Gamma (S \to \gamma\gamma) \exp [ -m_{S}/T_{\rm BBN}] > {\cal H} \,.
%\end{eqnarray}
%More details of the BBN limits on $m_S$ and $\sin\theta$ can be found e.g. in Ref.~\cite{Fradette:2017sdd}.
%\end{itemize}

%is in between the solar temperature and the WD temperature, i.e. $T_\oplus <u < T_{\rm WD}$, the solar limits  will be the same as in Fig.~\ref{fig:limits}, and all other stellar limits are weaken by a factor of $\sqrt{4\pi^2} = 2\pi$, and the WD limit still excludes the XENON1T excess region.

%In particular, if the VEV $u$ lies in between the Sun temperature of 1 keV and the white dwarf temperature of $T_{\rm WD} \simeq 6$ keV, i.e. $1 \; {\rm keV} < u <6$ keV, then when the temperature $T > u$ we have the electroweak VEV $v \neq 0$ and $u = 0$, and the $\phi$ field couples to the SM Higgs via the $\lambda_{H\phi}$ term in Eq.~(\ref{eqn:potential}) and does not mix with it.

\section{Conclusion}
\label{sec:conclusion}

Astrophysical objects like the Sun,  RGs, WDs and HB stars can be used to set limits on light BSM particles. In this paper we use the  observed luminosity limits from these stellar objects to constrain a light CP-even scalar mass and its couplings to the SM particles. For a generic scalar $S$ that mixes with the SM Higgs boson, the dominant production channel of $S$ in the stellar core is from the electron-nuclei bremsstrahlung process. As a result of the larger coupling of the SM Higgs to nucleons than to electrons and photons, the bremsstrahlung process is more important than the Compton-like and Primakoff-like processes (cf.~Fig.~\ref{fig:channels}). Taking into account the production, decay and reabsorption of $S$ inside the stars, we find that the stellar luminosity limits exclude the scalar mass up to 350 keV and the mixing angle $\sin\theta$  ranging from {$7.0\times10^{-18}$} to $3.4\times10^{-3}$, as presented in Fig.~\ref{fig:limits} and Table~\ref{tab:limits}. These stellar limits are more stringent than those from the supernovae, invisible meson decay and precision Higgs data for scalar mass $m_S \lesssim 200$ keV.

When one applies these stellar limits to specific models, there might be some differences, as illustrated in Section~\ref{sec:singlet}. Here we extend the SM Higgs sector by a $Z_2$-invariant  real singlet scalar.
%We have also considered also two classes of specific models, which are to some extent different from the generic case.
%The first class is the extension of the SM Higgs,
Then the $h-S$ mixing angle can be related to the parameters in the scalar potential in Eq.~(\ref{eqn:potential}), depending on the singlet scalar VEV $v_S$. The corresponding stellar limits are presented in Fig.~\ref{fig:limits:singlet}. The analysis in this paper can be extended to other scalar models, like a leptonic scalar~\cite{Dev:2017ftk}, where the couplings to nucleons are relatively small. However, in this case the plasma effect might be more important, for both production and reabsorption of $S$ in the stars~\cite{Hardy:2016kme}. The full analysis of plasma effects for the stellar limits on a leptonic scalar is beyond the main scope of this paper, and will be pursued in a future publication.

%The second class is the leptonic scalar model which can arise in some of the well-motivated BSM scenarios. In this case, the production of $S$ in the stars are dominated by the Compton-like and electron-nuclei bremsstrahlung processes, which are both induced by its coupling $h_{ee}$ to electrons. The stellar limits for the leptonic scalar are shown in Fig.~\ref{fig:limits:leptonic} \blue{and Table~\ref{tab:limits:leptonic}}, which exclude the coupling range $1.1 \times 10^{-17}\lesssim h_{ee} \lesssim 1.3 \times10^{-6}$.
%For completeness, we have calculated also the supernova limits on the leptonic scalar $S$, which exclude the coupling range of $6.3\times10^{-10} \lesssim h_{ee} \lesssim 6.4 \times 10^{-9}$ up to scalar mass of 12 MeV. As seen in Fig.~\ref{fig:limits:sn}, this is very different from the supernova limits in the generic scalar case,

We have also explored whether the light scalar under consideration can explain the recent XENON1T excess. Similar to the pseudoscalar axion case, a keV-scale light CP-even scalar can be abundantly produced in the Sun, and after reaching the Earth, it will be absorbed by electrons in the XENON1T detector, and thus can possibly be used to explain the XENON1T electron recoil excess. However, for both the generic scalar which mixes with the SM Higgs, as well as the real singlet scalar case studied here, the XENON1T excess parameter space has been excluded by the stringent stellar limits, as presented in Fig.~\ref{fig:excess}.

\section*{Acknowledgments}

We are grateful to Steven Harris for pointing out an error in the code for calculation of the mean free path of light scalars in the stars. {We also thank Edoardo Vitagliano for useful correspondence on the plasma effect}. The work of B.D. and Y.Z. is supported in part by the US Department of Energy under Grant No.  DE-SC0017987 and by a Fermilab Intensity Frontier Fellowship. This work was also supported in part by the Neutrino Theory Network Program under Grant No.  DE-AC02-07CH11359. The work of R.N.M. is supported by the US National Science Foundation grant no. PHY-1914631. Y.Z. is also partially supported by ``the Fundamental Research Funds for the Central Universities''.

\appendix
\section{Differential cross sections}
\label{sec:differential}

%For the sake of completeness we collect some of the differential cross sections in the appendix.
For the production of $S$ via the Compton-like process in Eq.~(\ref{eqn:Compton}), the differential cross section with respect to the $S$ energy $E_S$ is given by
\begin{eqnarray}
\label{eqn:Compton:differential}
\frac{d\sigma_{\rm C}}{dE_S} &=& \frac{\alpha y_e^2 \sin^2\theta}{8 m_e^3 y^2 }
\left[ -\frac{ 4 - 2y (2-x) - q^2 (1-y) }{2y^2}
\right. \nonumber \\
&& \left. \qquad  + \frac{ 2 ( 2 (2+x^2) + (y-2x) (3+y) ) }{y(2x-q^2)}
 - \frac{4(2-x)}{(2x-q^2)^2} \right] \, ,
\end{eqnarray}
where $q\equiv m_S/m_e$, $x \equiv E_S/m_e$ and $y\equiv E_\gamma/m_e$.
As in Eq.~(\ref{eqn:Compton:total}) we have neglected the kinetic energy of electrons in the initial state.
%In Eq.~(\ref{eqn:Compton:differential}), the function
%with $x \equiv = E_S/m_e$.
The range for $x$, which corresponds to the scalar energy $E_S$ in the final state, is
\begin{eqnarray}
\frac{ (1+y) (2y+q^2) - 2 y \tilde{y} }{2(1+2y)} < x <
\frac{ (1+y) (2y+q^2) + 2 y \tilde{y} }{2(1+2y)} \,,
\end{eqnarray}
where we have defined
\begin{eqnarray}
\tilde{y} \equiv \sqrt{\left[ y + q \left( 1-\frac12 q \right) \right] \left[ y - q \left( 1+\frac12 q \right) \right] } \,.
\label{eq:ytilde}
\end{eqnarray}
In the limit of a massless scalar, i.e. $m_S \to 0$, the differential cross section~\eqref{eqn:Compton:differential} can be simplified to
\begin{eqnarray}
\frac{d\sigma_{\rm C} (m_S \to 0)}{dE_S} &\simeq & \frac{\alpha y_e^2 \sin^2\theta}{8 m_e E_\gamma^2}
\left[\frac{ 4xy (y-x) - (2-xy) (x-y)^2 }{x^2y^2}\right] \,.
\end{eqnarray}

The function $f_{\rm C} (q,y)$ appearing in Eq.~\eqref{eqn:Compton:total} is given in terms of the $\tilde{y}$ defined in Eq.~\eqref{eq:ytilde} as
\begin{eqnarray}
\label{eqn:fC}
f_{\rm C} (q,y) &=& \frac{3}{16 y^3 (1+2y)^2}
\bigg\{ 2\tilde{y} \Big[ -2 (2+3y) (2+5y+y^2) + q^2 (2+8y+7y^2) \Big]  \nonumber \\
&& +(1+2y)^2  \Big( 2 (2+y)^2 -2q^2 (3+y) + q^4 \Big) \log \left( \frac{2(1+y +\tilde{y}) - q^2 }{2(1+y -\tilde{y}) - q^2 } \right) \bigg\} \,.
\end{eqnarray}
In the limit of $y \to 0$ and $q \to 0$, $f_{\rm C}(0,0)=1$.

%When the scalar energy is integrated out, the total cross section reads

\bibliographystyle{JHEP}
\bibliography{ref}

\end{document}